\documentclass[a4paper]{rnti}
\pdfoutput=1
\usepackage{graphicx}

\titrecourt{Un index de jointure pour les entrep\^{o}ts de donn\'{e}es XML}

\nomcourt{H. MAHBOUBI et al.}

\titre{Un index de jointure pour les entrep\^{o}ts de donn\'{e}es XML }

\auteur{Hadj Mahboubi, Kamel Aouiche, J\'{e}r\^{o}me Darmont}

\affiliation{
    ERIC, Universit\'{e} Lumi\`{e}re Lyon 2\\
    5 avenue Pierre Mend\`{e}s-France\\
    69676 Bron Cedex\\
    \{ hmahboubi | kaouiche | jdarmont\}@eric.univ-lyon2.fr
}

\resume{Les entrep\^{o}ts de donn\'{e}es XML proposent une base int\'{e}ressante pour les
applications d\'{e}cisionnelles qui exploitent des donn\'{e}es h\'{e}t\'{e}rog\`{e}nes et provenant
de sources multiples. Cependant, les performances des SGBD natifs XML \'{e}tant
actuellement limit\'{e}es, il est n\'{e}cessaire de trouver des moyens de les
optimiser. Dans cet article, nous proposons un nouvel index sp\'{e}cifiquement
adapt\'{e} \`{a} l'architecture multidimensionnelle des entrep\^{o}ts de donn\'{e}es XML, qui
\'{e}limine le co\^{u}t des jointures tout en pr\'{e}servant l'information contenue dans
l'entrep\^{o}t initial. Une \'{e}tude th\'{e}orique et des r\'{e}sultats exp\'{e}rimentaux
d\'{e}montrent l'efficacit\'{e} de notre index, m\^{e}me lorsque les requ\^{e}tes sont
complexes.}

\summary{XML data warehouses form an interesting basis for decision-support
applications that exploit heterogeneous data from multiple sources. However,
XML-native database systems currently bear limited performances and it is
necessary to research ways to optimize them. In this paper, we propose a new
index that is specifically adapted to the multidimensional architecture of XML
warehouses and eliminates join operations, while preserving the information
contained in the original warehouse. A theoretical study and experimental
results demonstrate the efficiency of our index, even when queries are
complex.}

\usepackage{ifpdf}
\ifpdf
\else

\fi

\begin{document}
\section{Introduction}
\label{sec:0}

Les technologies entrant en compte dans les processus d\'{e}cisionnels, comme les
entrep\^{o}ts de donn\'{e}es (\emph{data warehouses}), l'analyse multidimensionnelle en
ligne (\emph{On-Line Analysis Process} ou OLAP) et la fouille de donn\'{e}es
(\emph{data mining}), sont d\'{e}sormais tr\`{e}s efficaces pour traiter des donn\'{e}es
simples, num\'{e}riques ou symboliques. Cependant, les donn\'{e}es exploit\'{e}es dans le
cadre des processus d\'{e}cisionnels sont de plus en plus complexes. L'av\`{e}nement du
Web et la profusion de donn\'{e}es multim\'{e}dia ont en grande partie contribu\'{e} \`{a}
l'\'{e}mergence de cette nouvelle sorte de donn\'{e}es. Dans ce contexte, le langage
XML peut grandement aider \`{a} l'int\'{e}gration et \`{a} l'entreposage de ces donn\'{e}es.
C'est pourquoi nous nous int\'{e}ressons aux travaux \'{e}mergents sur les entrep\^{o}ts de
donn\'{e}es XML~\citep{XCube,DAWAX,P01,GRV01}. Cependant, les requ\^{e}tes
d\'{e}cisionnelles exprim\'{e}es en XML sont g\'{e}n\'{e}ralement complexes du fait qu'elles
impliquent de nombreuses jointures et agr\'{e}gations. Par ailleurs, les syst\`{e}mes
de gestion de bases de donn\'{e}es (SGBD) natifs XML pr\'{e}sentent actuellement des
performances m\'{e}diocres quand les volumes de donn\'{e}es sont importants ou que les
requ\^{e}tes sont complexes. Il est donc crucial lors de la construction d'un
entrep\^{o}t de donn\'{e}es XML de garantir la performance des requ\^{e}tes XQuery qui
l'exploiteront.

Plusieurs \'{e}tudes traitent de l'indexation des donn\'{e}es
XML~\citep{GSZ,LG02,CMS02}. Ces index optimisent principalement des requ\^{e}tes
exprim\'{e}es en expressions de chemin. Or, dans le contexte des entrep\^{o}ts de
donn\'{e}es XML, les requ\^{e}tes sont complexes et comportent plusieurs expressions de
chemin. De plus, ces index op\`{e}rent sur un seul document et ne prennent pas en
compte d'\'{e}ventuelles jointures, qui sont courantes dans les requ\^{e}tes
d\'{e}cisionnelles. \`{A} notre connaissance, seul l'index Fabric~\citep{Fabric} permet
actuellement g\'{e}rer plusieurs documents XML. Cependant, cet index ne prend pas
en compte les relations qui peuvent exister entre ces documents (les pr\'{e}dicats
de jointure, notamment) et n'est donc pas non plus adapt\'{e} \`{a} nos besoins.

C'est pourquoi nous proposons une structure d'index sp\'{e}cifiquement adapt\'{e}e aux
donn\'{e}es multidimensionnelles d'un entrep\^{o}t XML, c'est-\`{a}-dire une structure
capable de maintenir des donn\'{e}es de plusieurs documents XML \`{a} la fois tout en
pr\'{e}servant l'information contenue dans ces donn\'{e}es et leur mod\'{e}lisation en
\'{e}toile. Notre structure d'index, que nous qualifions d'index de jointure,
permet \'{e}galement d'assurer un meilleur traitement des requ\^{e}tes d\'{e}cisionnelles
exprim\'{e}es en XQuery en \'{e}liminant les co\^{u}ts de jointure. Afin de valider cette
proposition, nous avons men\'{e} une \'{e}tude th\'{e}orique ainsi que des exp\'{e}rimentations
sur un entrep\^{o}t de donn\'{e}es XML r\'{e}el.

Le reste de cet article est organis\'{e} comme suit. Nous pr\'{e}sentons dans la
section~\ref{sec:1} le contexte de notre \'{e}tude ainsi que notre structure
d'index. Les \'{e}tudes th\'{e}orique et exp\'{e}rimentale que avons men\'{e}es pour tester la
validit\'{e} de notre index sont pr\'{e}sent\'{e}s dans la section~\ref{sec:2}. Finalement,
nous concluons et discutons nos perspectives de recherche dans la
section~\ref{sec:3}.

\section{Indexation des entrep\^{o}ts de donn\'{e}es XML}
\label{sec:1}

\subsection{Contexte}
\label{sec:1.1}

Afin d'appliquer notre d\'{e}marche d'optimisation des performances, nous avons
s\'{e}lectionn\'{e} l'architecture d'entrep\^{o}t de donn\'{e}es XCube~\citep{XCube}, qui
propose une mod\'{e}lisation en \'{e}toile de donn\'{e}es stock\'{e}es dans des documents XML.
Ces documents permettent de repr\'{e}senter respectivement le sch\'{e}ma et les
m\'{e}tadonn\'{e}es (\emph{Schema.xml)}, les dimensions (\textit{Dimensions.xml}) et la
table de faits (\textit{TableFacts.xml}) de l'entrep\^{o}t XML. La figure
\ref{dimfact} (a) et \ref{dimfact} (b) repr\'{e}sentent les deux derniers documents
sous forme de graphes XML.

Pour l'interrogation de l'entrep\^{o}t, nous avons adopt\'{e} le langage XQuery car il
permet de formuler des requ\^{e}tes complexes. La requ\^{e}te \textbf{Q} donne un
exemple d'une requ\^{e}te d\'{e}cisionnelle exprim\'{e}e dans ce langage. Cette requ\^{e}te
retourne la moyenne des quantit\'{e}s vendues aux clients de Lyon, avec
regroupement par nom et codes postal. Elle r\'{e}alise une jointure entre le
document \textit{Dimensions.xml} et \textit{TableFacts.xml}. Notons que XQuery
ne permet pas normalement de faire des op\'{e}rations de groupement \textit{Group
by} multiples. Dans notre impl\'{e}mentation, nous avons donc \'{e}tendu l'interface
d'interrogation du SGBD natif XML eXist~\citep{M02} de mani\`{e}re \`{a} ce qu'il
puisse traiter ce type de requ\^{e}tes.

\textbf{Q :}
\begin{footnotesize}
\textit{\textsf{ \textbf{for} \$a \textbf{in}
//dimensionData/classification/Level[@node='customers']/node, \$x
\textbf{in}//CubeFacts/cube/Cell\\
\textbf{where} \$b/attribute[@name=`cust\_city',@value=`Lyon']  \textbf{and}
\$x/dimension /@node=\$a/@id  \textbf{and} \$x/dimension\\/@id=`customers'\\
\textbf{group by}(@cust\_first\_name,@cust\_postal\_code)  \textbf{return}
\textbf{sum}(quantity)}}
\end{footnotesize}

\subsection{Structure de notre index de jointure}

%Construire les structures d'index existantes sur un entrep\^{o}t de donn\'{e}es
%mod\'{e}lis\'{e} selon la sp\'{e}cification XCube provoque une perte d'information pour la
%r\'{e}solution des requ\^{e}tes de jointure en raison de la disparition des relations
%entre mesures et dimensions. Nous proposons donc de construire une structure
%d'index qui permet de conserver ses relation, c'est \`{a} dire les relations entre
%dimensions et mesures dans un fait.

%\subsubsection{Structure de notre index de jointure}

Notre index doit permettre de conserver les relations entre les dimensions et
les mesures dans un fait. Il pr\'{e}sente donc une structure similaire \`{a} celle du
document \textit{TableFacts.xml}, \`{a} l'exception de l'\'{e}l\'{e}ment
\textit{attribute}.

\begin{figure}[h]
{\centering
\resizebox*{1\textwidth}{!}{\includegraphics{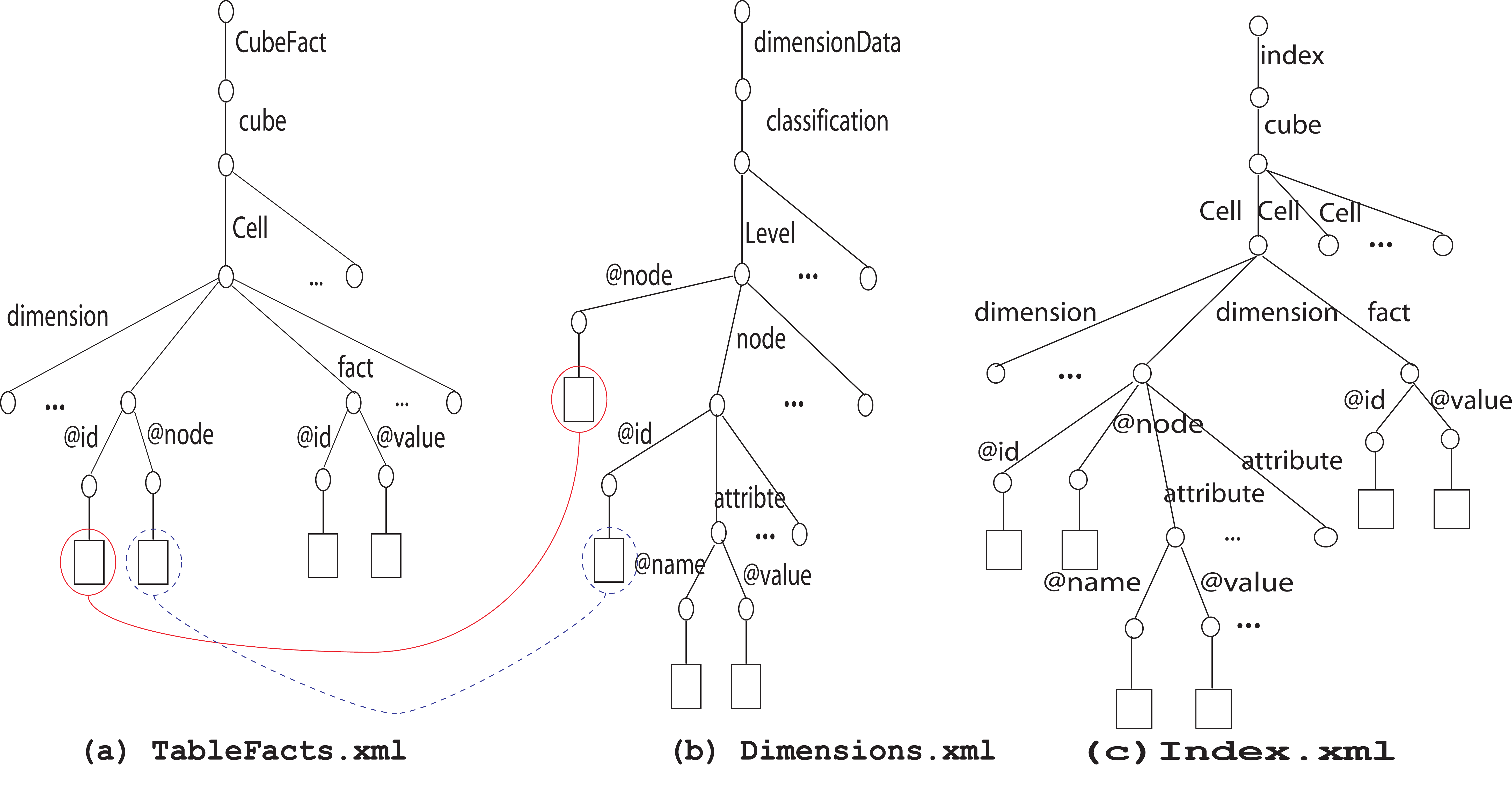}}
\par}
\caption{Structure en arbre des documents (a) \textit{Dimensions.xml}, (b)
\textit{TableFacts.xml} et (c) \textit{Index.xml}} \label{dimfact}
\end{figure}

Sa structure est pr\'{e}sent\'{e}e dans la figure~\ref{dimfact} (c). Les \'{e}tiquettes qui
commencent par le caract\`{e}re $@$ repr\'{e}sentent les attributs et les autres
repr\'{e}sentent les \'{e}l\'{e}ments. Chaque cellule est identifi\'{e}e par des dimensions et
un ou plusieurs faits. Un fait (\'{e}l\'{e}ment \textit{Fact}) poss\`{e}de deux attributs,
$@id$ et $@value$, qui indiquent respectivement son nom et sa valeur. Chaque
dimension (\'{e}l\'{e}ment \textit{dimension}) est identifi\'{e}e par deux attributs: $@id$
qui donne le nom de la dimension et $@node$ qui donne la valeur de
l'identifiant de la dimension. De plus, l'\'{e}l\'{e}ment dimension poss\`{e}de un certain
nombre d'\'{e}l\'{e}ments fils (\'{e}l\'{e}ments \textit{attribute}). Ces \'{e}l\'{e}ments sont ins\'{e}r\'{e}s
pour stocker les noms et les valeurs des attributs de chaque dimension. Ils
sont obtenus depuis le document \textit{Dimensions.xml}. Un \'{e}l\'{e}ment
\textit{attribute} est caract\'{e}ris\'{e} par deux attributs, $@nom$ et $@value$, qui
indiquent respectivement le nom et la valeur de chaque attribut.

La migration des donn\'{e}es des documents \textit{Dimensions.xml} et
\textit{TableFacts.xml} vers la structure d'index, et le fait de stocker dans
une m\^{e}me cellule les faits, les dimensions et leurs attributs, nous permet
d'\'{e}liminer les op\'{e}rations de jointure. Toutes les informations n\'{e}cessaires pour
la jointure sont en effet stock\'{e}es dans la m\^{e}me cellule.

%Le nombre
%d'\'{e}l\'{e}ments \textit{attribute} d\'{e}pend du nombre d'attributs de chaque dimension.

%\begin{figure}[h]
%{\centering \resizebox*{0.4\textwidth}{!}{\includegraphics{req_exp.eps}}
%\par}
%\caption{Structure de notre index de jointure} \label{index}
%\end{figure}

%\begin{figure}[t]
% \begin{minipage}[b]{.5\linewidth}
%  \centering\epsfig{figure=index_xml.eps,width=\linewidth}
%  \caption{Structure de notre index de jointure \label{index}}
% \end{minipage} \hfill
% \begin{minipage}[b]{.5\linewidth}
%  \centering\epsfig{figure=req.eps,width=\linewidth}
%  \caption{Exemple de requ\^{e}te d\'{e}cisionnelle \label{req}}
% \end{minipage}
%\end{figure}
\section{Validation}
\label{sec:2}

\subsection{\'{E}tude th\'{e}orique}

Une requ\^{e}te type d\'{e}finie sur un entrep\^{o}t de donn\'{e}es XML mod\'{e}lis\'{e} selon la
sp\'{e}cification XCube r\'{e}alise plusieurs jointures entre les faits stock\'{e}s dans
\textit{TableFacts.xml} et les dimensions de \textit{Dimensions.xml}. Il faut
alors v\'{e}rifier les contraintes suivantes : $TableFacts.@id=Dimensions.@node$ et
$TableFacts.@node = Dimensions.@id$. La premi\`{e}re \'{e}galit\'{e} v\'{e}rifie que la
dimension composant une cellule est bel et bien la dimension exprim\'{e}e dans la
requ\^{e}te. La seconde v\'{e}rifie que le n\oe ud d'une dimension (\'{e}quivalent d'une
cl\'{e} primaire) correspond (peut \^{e}tre joint) au n\oe ud, de la m\^{e}me dimension,
d\'{e}fini dans une cellule (\'{e}quivalent d'une cl\'{e} \'{e}trang\`{e}re de la table de faits).

L'ex\'{e}cution d'une requ\^{e}te sans utilisation de notre index peut se d\'{e}rouler
comme suit. Pour chaque dimension d\'{e}finie par \textit{@node='nom de la
dimension'}, les identifiants $@id$ v\'{e}rifiant la clause \textit{Where} sont
recherch\'{e}s. Le document \textit{Dimensions.xml} est parcouru en profondeur,
jusqu'au n\oe ud \textit{Level}. Les fils $node$ du n\oe ud $Level$ sont
ensuite parcourus en largeur jusqu'\`{a} trouver le n\oe ud dont la valeur de
$@node$ est \'{e}gale au nom de la dimension sp\'{e}cifi\'{e} dans la requ\^{e}te. Le co\^{u}t de
ce parcours est \'{e}gal au nombre de n\oe uds $Level$ du document
\textit{Dimensions.xml}, c'est-\`{a}-dire le nombre de dimensions dans le sch\'{e}ma,
d\'{e}not\'{e} $|dimension|$. Si plusieurs dimensions sont d\'{e}finies, le parcours donne
alors lieu \`{a} autant de n\oe uds que de dimensions. En revanche, le co\^{u}t de ce
parcours reste invariant car tous les n\oe uds $Level$ sont parcourus. Pour
chaque n\oe ud trouv\'{e}, ses fils sont parcourus en profondeur jusqu'\`{a} trouver la
liste des $@id$ v\'{e}rifiant les conditions \textit{@name='nom de l'attribut'} et
\textit{@value='valeur de l'attribut'}. Le co\^{u}t de ce parcours est \'{e}gal au
nombre de fils $attribute$. En r\'{e}sum\'{e}, le co\^{u}t de traitement des dimensions est
\'{e}gal \`{a} $|a_{i}|*|d_{i}|$, o\`{u} $|a_{i}|$ d\'{e}signe le nombre d'attributs de chaque
dimension et $|d_{i}|$ le nombre d'\'{e}l\'{e}ments $node$, c'est-\`{a}-dire le nombre de
fils de chaque dimension. Pour r\'{e}aliser une jointure entre les dimensions du
document \textit{Dimensions.xml} et les faits du document
\textit{TableFacts.xml}, les $@id$ retrouv\'{e}s dans le traitement des dimensions
sont recherch\'{e}s dans les faits.  Le document \textit{TableFacts.xml} est alors
parcouru en profondeur jusqu'au niveau $Cell$. Le co\^{u}t de ce parcours est \'{e}gal
\`{a} 2. Les cellules sont ensuite parcourues en largeur afin de trouver les
dimensions dont le fils $@id$ est \'{e}gal \`{a} $@node$ de \textit{Dimensions.xml} et
$@node$ est \'{e}gal \`{a} $@id$ de \textit{Dimensions.xml}. En r\'{e}sum\'{e}, le co\^{u}t de
traitement du document \textit{TableFacts.xml} est $|Cell|$, o\`{u} $|Cell|$ est le
nombre de cellules du document \textit{TableFacts.xml}. Le co\^{u}t de traitement
d'une requ\^{e}te est donn\'{e}e par la formule
$E_{sans-index}=((|Cell|)*|Dimension|)*(|Dimension|+(|d_{i}|*|a_{i}|))$.

%~\ref{eq1}.

%\begin{eqnarray}
%E_{sans-index}=((|Cell|)*|Dimension|)*(|Dimension|+(|d_{i}|*|a_{i}|))
%\label{eq1}
%\end{eqnarray}

L'ex\'{e}cution d'une requ\^{e}te, avec utilisation de notre index (stock\'{e} dans le
document \textit{Index.xml}), peut se d\'{e}rouler comme suit. Pour chaque
dimension d\'{e}finie par \textit{@node='nom de la dimension'}, les identifiants
$@id$ v\'{e}rifiant la clause \textit{Where} sont recherch\'{e}s. Le document
\textit{Index.xml} est parcouru en profondeur, jusqu'au n\oe ud \textit{Cell}.
Le co\^{u}t de ce parcours est \'{e}gal au nombre de cellules dans le document
\textit{Index.xml}. Les fils $dimension$ du n\oe ud $Cell$ sont ensuite
parcourus en largeur jusqu'\`{a} trouver le n\oe ud dont la valeur de $@id$ est
\'{e}gale au nom de la dimension sp\'{e}cifi\'{e} dans la requ\^{e}te. Le co\^{u}t de ce parcours
est \'{e}gal au nombre de n\oe uds $dimension$ du document \textit{Index.xml},
c'est-\`{a}-dire le nombre de dimensions dans le sch\'{e}ma de l'entrep\^{o}t de donn\'{e}es :
$|dimension|$. Pour chaque n\oe ud trouv\'{e}, ses fils sont parcourus en
profondeur jusqu'\`{a} trouver le n\oe ud $attribute$ v\'{e}rifiant les conditions
\textit{@name='nom de l'attribut'} et \textit{@value='valeur de l'attribut'}.
Le co\^{u}t de ce parcours est \'{e}gal au nombre de fils $attribute$
 $|a_{i}|$. En r\'{e}sum\'{e}, le co\^{u}t de traitement d'une requ\^{e}te qui
exploite notre structure d'index est donn\'{e} par la formule
$E_{index}=|Cell|*(|Dimension|+|a_{i}|) \label{eq2}$.
%~\ref{eq2}.

%\begin{eqnarray}
%E_{index}=|Cell|*(|Dimension|+|a_{i}|) \label{eq2}
%\end{eqnarray}

La figure~\ref{res} (a) repr\'{e}sente la variation des co\^{u}ts $E_{sans-index}$ et
$E_{index}$ en fonction du nombre de cellules. Nous constatons que
l'utilisation de notre index permet un gain de facteur 14000 en moyenne.

\subsection{Exp\'{e}rimentations}

\begin{figure}[h]
{\centering \resizebox*{1\textwidth}{!}{\includegraphics{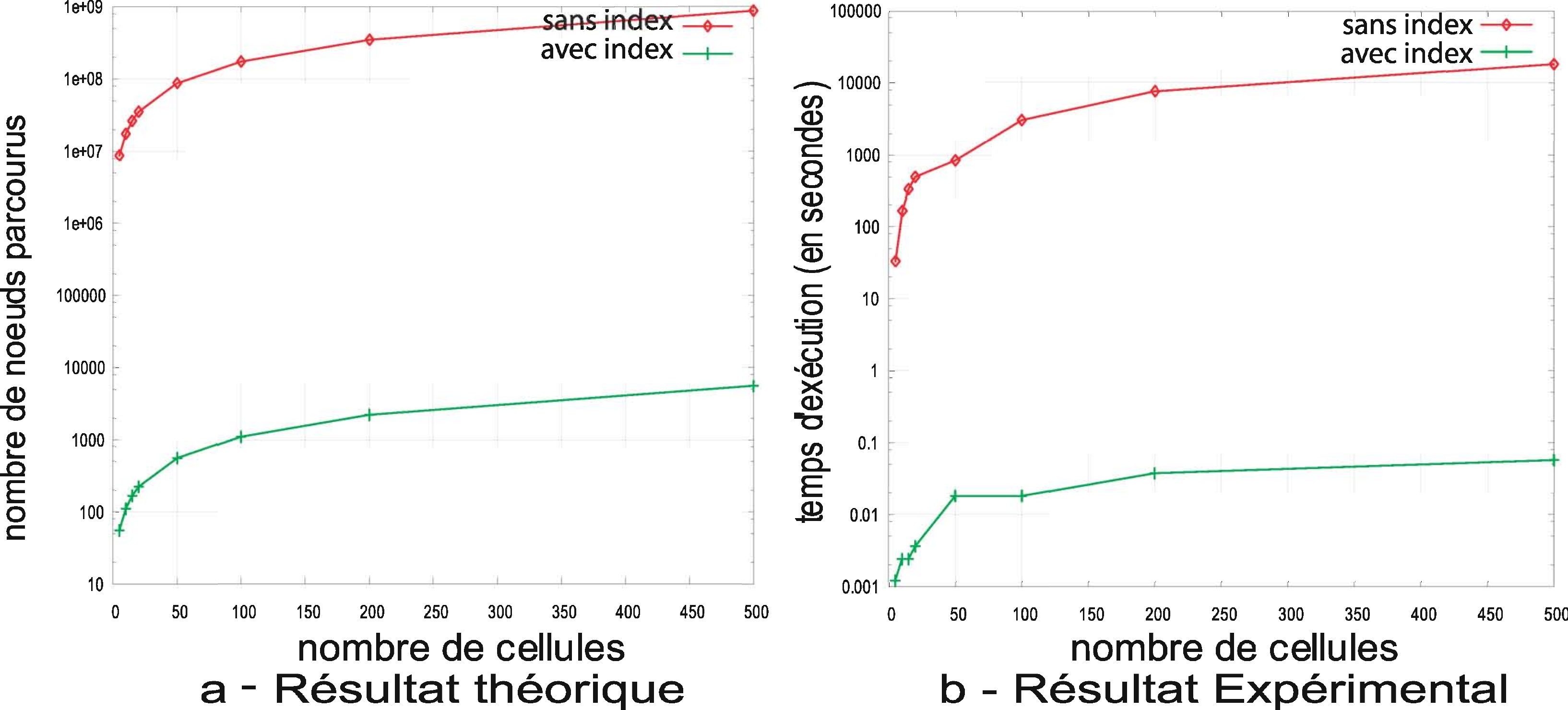}}
\par}
   \caption{R\'{e}sultats}
   \label{res}
\end{figure}

En compl\'{e}ment de notre \'{e}tude th\'{e}orique, nous avons effectu\'{e} des
exp\'{e}rimentations afin de tester l'efficacit\'{e} de notre proposition de structure
d'index. Nous avons g\'{e}n\'{e}r\'{e} un entrep\^{o}t de donn\'{e}es XCube implant\'{e} au sein du
SGBD XML natif eXist. Construit \`{a} partir d'un entrep\^{o}t de donn\'{e}es relationnel
test classique, cet entrep\^{o}t est constitu\'{e} d'une table de faits \textit{sales}
(4,92~Mo) et de cinq dimensions \textit{channels}, \textit{promotions},
\textit{customers}, \textit{products} et \textit{times} (3,77~Mo). Nous avons
effectu\'{e} nos tests sur une machine dot\'{e}e d'un processeur Intel Pentium 4 GHz
avec 1~GB de m\'{e}moire et un disque dur IDE. Nous avons ex\'{e}cut\'{e} la requ\^{e}te
d\'{e}cisionnelle XQuery de la section~\ref{sec:1.1} avec et sans utilisation de
notre index et en faisant varier la taille de l'entrep\^{o}t. La figure~\ref{res}
(b) pr\'{e}sente les r\'{e}sultats obtenus exprim\'{e}s en temps d'ex\'{e}cution par rapport au
nombre de cellules (faits) dans le document \textit{TableFacts.xml}.

La figure~\ref{res} (b) montre qu'avec l'utilisation de notre index, nous
obtenons des temps de traitement en moyenne 25669 fois inf\'{e}rieurs \`{a} ceux
obtenus sans utiliser notre index. De plus, cette figure est semblable \`{a} celle
qui repr\'{e}sente nos estimations th\'{e}oriques (figure~\ref{res} (b)). Nous avons
\'{e}galement utilis\'{e} notre structure d'index pour traiter la requ\^{e}te sur la
totalit\'{e} des cellules du documents \textit{TableFatcs.xml}. Nous avons obtenu
un temps d'ex\'{e}cution de moins de deux secondes, alors que le syst\`{e}me s'est
av\'{e}r\'{e} incapable de r\'{e}pondre \`{a} la requ\^{e}te dans un temps raisonnable lorsque nous
n'utilisions pas notre index.

\section{Conclusion et perspectives}
\label{sec:3} Nous avons pr\'{e}sent\'{e} dans cet article un nouvel index de jointure
sp\'{e}cifiquement adapt\'{e} aux entrep\^{o}ts de donn\'{e}es XML. Cette structure de donn\'{e}es
permet d'optimiser les temps d'acc\`{e}s \`{a} plusieurs documents XML en \'{e}liminant le
co\^{u}t de jointure, tout en pr\'{e}servant l'information contenue dans l'entrep\^{o}t
initial. Une \'{e}tude de complexit\'{e} et des exp\'{e}rimentations nous ont permis de
d\'{e}montrer l'efficacit\'{e} de notre index m\^{e}me lorsque les requ\^{e}tes sont complexes
et la taille de l'entrep\^{o}t volumineuse. Ces exp\'{e}rimentations nous ont par
ailleurs amen\'{e}s \`{a} \'{e}tendre la syntaxe des requ\^{e}tes XQuery afin qu'elles puissent
supporter des clauses de regroupement (\textit{Group by}) multiples.

Les perspectives de ce travail se situent dans le cadre du d\'{e}veloppement des
SGBD natifs XML, qui vise \`{a} les doter des m\^{e}mes fonctionalit\'{e}s et performances
que les SGBD relationnels. Premi\`{e}rement, notre index pourrait \^{e}tre directement
int\'{e}gr\'{e} au sein d'un SGBD natif XML. Il serait \'{e}galement indispensable de
mettre au point une strat\'{e}gie de maintenance incr\'{e}mentale de la structure de
donn\'{e}es de notre index lorsque les donn\'{e}es sources sont mises \`{a} jour. Par
ailleurs, notre m\'{e}canisme de r\'{e}\'{e}criture de requ\^{e}tes pourrait \'{e}galement \^{e}tre
directement int\'{e}gr\'{e}, notamment au niveau de la mod\'{e}lisation des vues.
Finalement, les modifications que nous avons apport\'{e}es \`{a} la clause de
regroupement \textit{Group by} de XQuery pourraient aussi faire objet d'une
proposition d'extension de la syntaxe de ce langage.

\bibliographystyle{rnti}

\bibliography{egc2006}

\end{document}